\documentclass{modified}

\begin{document}

\title{GAUCHE-TRANS ENERGY DIFFERENCES\\
IN DIMETHOXYMETHANE AND DIMETHOXYETHANE
}

\author{\footnotesize E.V.R.CHAN}
 
\maketitle

\begin{abstract}
	Ab-initio self consistent field calculations using double zeta
Gaussian basis set expansions were performed on conformers of
Dimethoxymethane and Dimethoxyethane. The gauche-trans energy differences
using the rigid rotor approximation were calculated.
\end{abstract}

\section{DMM}

	Four conformers are possible for DMM (Dimethoxymethane also known as
dioxymethylene dimethylether). They are 
TT (delta1 = delta2 = 18$0^\circ$, E = energy =  
-267.796016 a.u.) , TG (delta1 = 18$0^\circ$, delta2 = -6$0^\circ$, E = -267.802131 a.u.),
GG (delta1 = delta2 = -63.$3^\circ$, E = -267.808081 a.u.)  and G-G+
(delta1 = -63.$3^\circ$ , delta2 = +56.$7^\circ$ , E = -267.737630 a.u.).  See figure 1
for definitions of the dihedral angles delta1 and delta2.

\begin{figure}[th]
\centerline{\psfig{file=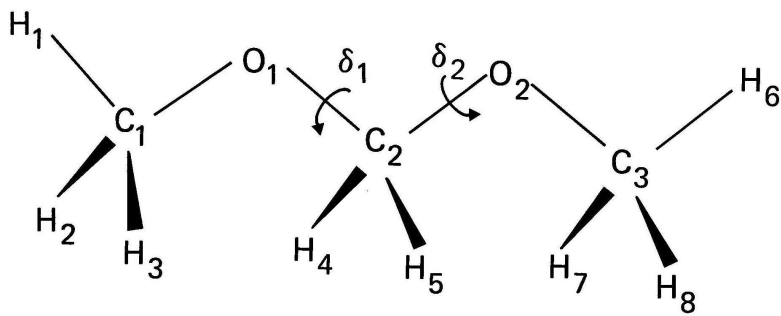,width=11cm}}
\vspace*{8pt}
\caption{ Dimethoxymethane } 
\end{figure}

A dihedral angle of close to ±l8$O^\circ$ corresponds to trans 
or T, in the vicinity of ±6$0^\circ$ to gauche or G. 
Ab-initio SCF ( self-consistent field ) calculations used 
a double-zeta basis set (Appendix A), obtained
from the atomic base$s^1$
contracted according to the method systematically
studied by Dunnin$g^2$. 
Among the four conformers, GG (delta1 = delta2 = -63.$3^\circ$ ) was the
most stable (E = -267.808081 a.u.). The geometr$y^3$
used was all tetrahedral angles, C-H = 1.10 \AA,
C-O = 1.43 \AA, all other dihedral angles are 18$0^\circ$ .  Fixing
delta1 = -63.$3^\circ$ and varying delta2 (the dihedral angle around the
C2 - O2 bond) gives a value of the gauche-trans energy difference in the
rigid rotor approximatio$n^4$.
The variation of energy with dihedral angle is
shown in Figure 2. 

\begin{figure}[th]
\centerline{\psfig{file=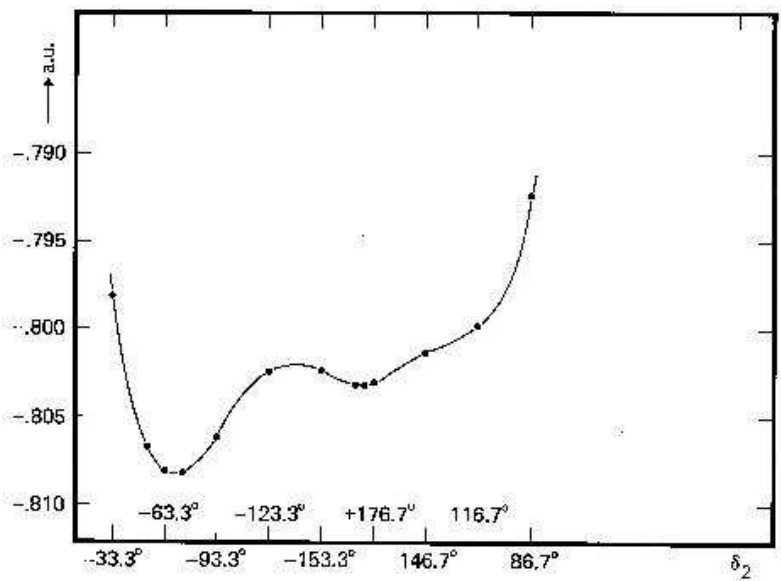,width=11cm}}
\vspace*{8pt}
\caption{ Dimethoxymethane delta1 = -63.$3^\circ$ }
\end{figure}

The trans minimum is very
shallow; the barrier for the T $\rightarrow$ G transition 
is about .6 Kcal/mole or the
order of RT at room temperature. A parabola was fitted to the T and G-
regions and the minimum energies were extrapolated to be
-267.8030934 au. ( -l76.8$7^\circ$ ) and -267.8083169 au. ( -69.5$5^\circ$ )
This corresponds to 3.3 Kcal/mole as the energy difference between the
gauche and trans minima (using lau. = 627.3047712 Kcal/mole).  These 
are in agreement with ab-initio, forcefields
and experimental studies cited  by Li$i^5$ etal. 
Steric repulsion of the hydrogens in the G-G+ position is very strong (greater
than 50 Kcal/mole above absolute minimum); the expected third minimum near
+63.$3^\circ$ is not obvious with this mesh size.  

\section{DME}

      For Dimethoxyethane (also known as dioxyethylene dimethylether and
abbreviated here as DME) energies of the four lowest conformers 
were calculated . They are TTT (18$0^\circ$, 18$0^\circ$, 18$0^\circ$, 
E = -306.815493 a.u.), 
TGT (18$0^\circ$, -6$0^\circ$, 18$0^\circ$, E = -306.811580 a.u.), 
TTG (18$0^\circ$, 18$0^\circ$, -6$0^\circ$, E = -306.803115 a.u.),  
 and  TGG (18$0^\circ$, -6$0^\circ$, -6$0^\circ$, E = -306.800219 a.u.).
See Figure 3 for definitions of dihedral angles.

\begin{figure}[th]
\centerline{\psfig{file=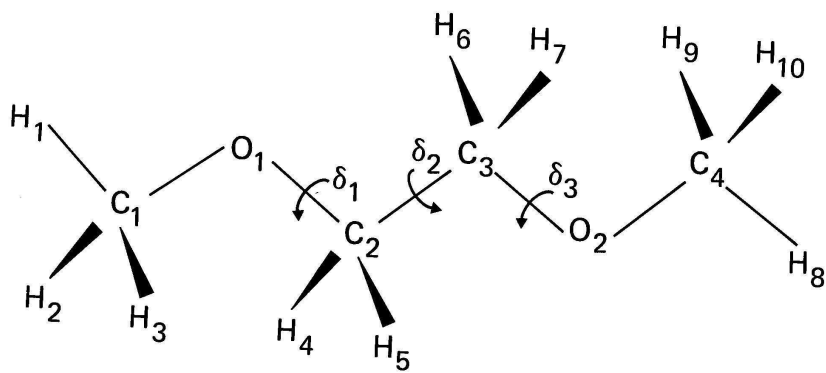,width=11cm}}
\vspace*{8pt}
\caption{ Dimethoxyethane }
\end{figure}

The double zeta basis set of Appendix A
was used. The geometry used was: all tetrahedral angles, C-C = l.53 \AA,
C-0 = l.43 \AA, C-H = l.10 \AA. Delta1 and delta3 were fixed at 18$0^\circ$  
and delta2 varied.  The data are displayed in Figure 4; because of
symmetry only half the values are shown. 

\begin{figure}[th]
\centerline{\psfig{file=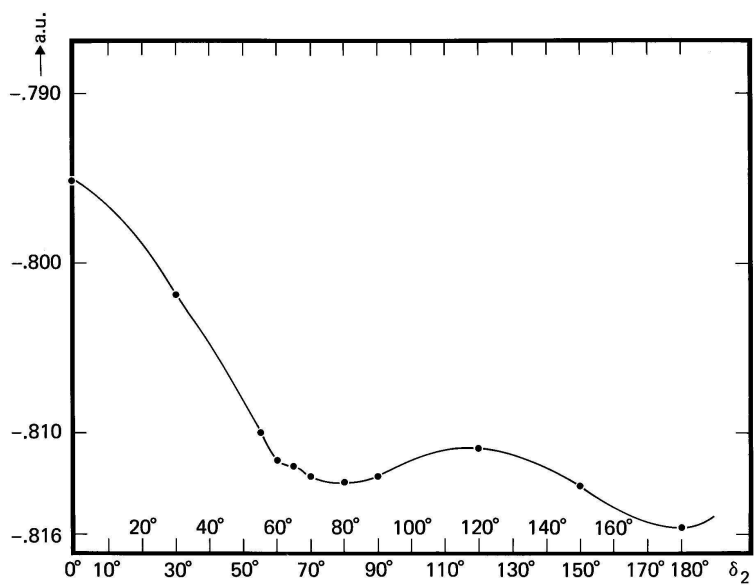,width=11cm}}
\vspace*{8pt}
\caption{ Dimethoxyethane delta1 = delta3 = 18$0^\circ$ }
\end{figure}

The secondary G minimum was
extrapolated by parabolic fit to   
-306.812966 au. ( 80.2$8^\circ$ ).  According to
Figure 4, there may be a shallow minimum around 6$0^\circ$  having a still higher
energy; however, the minimum at 80.2$8^\circ$  is clearly a lower energy and called
gauche (G). The trans-gauche energy difference is 1.6 Kcal/mole. The
G $\rightarrow$ T barrier is approximately 1.3 Kcal/mole (from Figure 4). 
These results are
in agreement with the calculations of Anderso$n^6$ etal.
and with the experimental
studies cited therein.  The basis set superposition error (BSSE) was 
calculated by Ha$n^7$ etal. 
to be .4 Kcal/mole and the gauche trans energy difference 1.4 - 1.5 Kcal./mole  
The repulsion (approximately 12.7 Kcal/mole 
above the trans minimum) in the delta2 = $0^\circ$ 
position is explained as due to interaction of oxygen lone pairs.

\section{ Appendix A }

\begin{verbatim}
Double-Zeta Basis
               EXPONENT      COEFFICIENT        

OXYGEN
S     .10662284940D+5     .79900000D-3    
S      .1599709689D+4     .61530000D-2 
S       .364725257D+3     .31157000D-1 
S       .103651793D+3     .11559600D+0
S        .33905805D+2     .30155200D+0
S        .12287469D+2     .44487000D+0
 * * * * * * * * * * * * * * * * * * * * *
S        .47568030D+1     .10000000D+1
 * * * * * * * * * * * * * * * * * * * * *
S        .10042710D+1	  .10000000D+1
 * * * * * * * * * * * * * * * * * * * * *
S        .30068600D+0     .10000000D+1
 * * * * * * * * * * * * * * * * * * * * * 
P        .34856463D+2     .15648000D-1
P        .78431310D+1     .98197000D-1
P        .23082690D+1     .30774800D+0 
P        .72316400D+0     .49247000D+0
 * * * * * * * * * * * * * * * * * * * * *
P        .21488200D+0     .10000000D+1
 * * * * * * * * * * * * * * * * * * * * * 
CARBON
S      .5240635258D+4     .93700000D-3
S       .782204795D+3     .72280000D-3
S       .178350830D+3     .36344000D-1
S        .50815942D+2     .13060000D+0
S        .16823562D+2     .31893100D+0
S        .61757760D+1     .43874200D+0
 * * * * * * * * * * * * * * * * * * * * *
S        .24180490D+1	  .10000000D+1
 * * * * * * * * * * * * * * * * * * * * *
S        .51190000D+0     .10000000D+1
 * * * * * * * * * * * * * * * * * * * * *
S        .15659000D+0     .10000000D+1
 * * * * * * * * * * * * * * * * * * * * *
P	 .18841800D+2     .13887000D-1
P        .41592400D+1     .86279000D-1     
P        .12067100D+1     .28874400D+0    
P        .38554000D+0     .49941100D+0
 * * * * * * * * * * * * * * * * * * * * *
P        .12194000D+0     .10000000D+1
 * * * * * * * * * * * * * * * * * * * * *
HYDROGEN
S	.188614448D+3     .71800000D-3
S        .28276596D+2     .55610000D-2
S        .64248300D+1     .28453000D-1
S        .18150410D+1     .10938100D+0
S        .59106300D+0     .30105700D+0
S        .21214900D+0     .47252200D+0
 * * * * * * * * * * * * * * * * * * * * *
S        .79891000D-1	  .10000000D+1
 * * * * * * * * * * * * * * * * * * * * *

ENDBASIS

\end{verbatim}

                     http://www.geocities.com/evrwebpage
\end{document}